\documentclass[superscriptaddress,showpacs,preprint]{revtex4}

\begin{document}

\title{Weak Measurement of Berry's Phase}

\author{Z. Gedik}

\email[]{gedik@sabanciuniv.edu}

\affiliation{Faculty of Engineering and Natural Sciences, Sabanci
University, Tuzla, Istanbul 34956, Turkey}

\date{\today}

\begin{abstract}
Quantum measurements can be generalized to include complex
quantities. It is possible to relate the quantum weak values of
projection operators to the third order Bargmann invariants. The
argument of the weak value becomes, up to a sign, equal to the
Berry's phase associated with the three state vectors. In case of
symmetric informationally complete, positive operator valued
measures (SIC-POVMs),  this relation takes a particularly simple
form. Alternating strong and weak measurements can be used to
determine Berry's phase directly, which demonstrates that not only
their real and imaginary parts but also moduli and arguments of weak
values have a physical significance. For an arbitrary projection
operator, weak value is real when the projector, pre- and
post-selected states lie on a so-called null phase curve which
includes the geodesic containing the three states as a special case.
\end{abstract}



\maketitle

Quantum weak values have been first introduced by Aharonov, Albert,
and Vaidman in the context of weak measurements \cite{Aharonov}. It
has been argued that the process of obtaining a weak value might not
be called as a quantum measurement \cite{Leggett,Peres,Reply}.
Independent of discussions on the relation between weak measurement
and notion of standard measuring procedure, weak values are well
defined quantities which can be beyond the range of eigenvalues
(greater than the maximum or smaller than the minimum)and they can
be complex \cite{Vaidman}.

Unitary evolution implied by the Schr\"{o}dinger equation  allows us
to evaluate not only the future quantum state vectors but also the
past ones. In other words, forward and backward evolutions are on
equal footing. On the other hand, because of the collapse
phenomenon, measurement is an irreversible process where there is no
time symmetry. The collapse destroys the information about the
initial state. Aharonov, Bergmann, and Lebowitz developed a theory
to formulate quantum process of measurement in a time symmetric way
\cite{Aharonov2}. For this purpose, they introduced two state
vectors instead of a single one. Having an initial state (or
so-called pre-selected state), we can obtain an ensemble subset by
choosing a particular final (so-called post-selected) state.
Additional information coming from the post-selected state leads to
time symmetry similar to the unitary evolution case. Physics of the
system during the time interval in-between can be described by a
forward evolving pre-selected and a backward evolving post-selected
states. That is why time-symmetric formulation is also know as the
two-state vector formalism of quantum mechanics.

In this work, we present an interpretation of weak values where
instead of real and imaginary parts, their moduli and arguments are
considered. It turns out that, the argument of the weak value can be
related to Berry's phase \cite{Berry}. More precisely, for
measurement of projection operators, the argument becomes nothing
but the phase of a third order Bargmann invariant \cite{Bargmann}.
Symmetric informationally complete, positive operator valued
measures (SIC-POVMs) can used to demonstrate experimentally the
suggested measurement method \cite{Jauch,Renes,Zauner}. In case of
projection operators, we show that weak value is real if the
projector, pre- and post-selected states are on a so-called null
phase curve where the geodesic connecting the two states is a
special case \cite{Rabei}.

The concept of weak values has been invented to obtain information
about the system between pre- and post-selected states. It is
possible to use the von Neumann model of measurement in the weak
interaction limit \cite{vonNeumann}. The system and the measurement
apparatus become in contact to obtain some information about the
system and yet the post-selected state is preserved \cite{Aharonov}.
The basic idea of weak measurement is to induce a weak coupling
between a system observable and a quantum pointer so that the time
evolution of the state vector is minimally modified. The coupling
can be made weak enough to  have the same outcome probabilities of
the final state relative to the no coupling situation and yet an
information about the system between initial $\mid\Psi\rangle$ and
final $\mid\Phi\rangle$ states can be obtained characterized by a
quantity called weak value \cite{Aharonov}
\begin{equation}
A_w=\frac{\langle\Phi\mid A\mid
\Psi\rangle}{\langle\Phi\mid\Psi\rangle}.
\end{equation}
If the overlap $\langle\Phi\mid\Psi\rangle$ between the pre- and
post-selected states is small enough, the weak value can be
arbitrarily large, beyond the maximum eigenvalue of $A$.
Furthermore, $A_w$ can be a complex number. Derivation of weak value
expression also gives hint on possible interpretation of complex
values. It has been argued that the imaginary part of the weak value
does not provide information about the observable but instead it
describes how the initial state would be unitarily disturbed
\cite{Dressel}.

According to the measurement axiom, measurement of a dynamical
variable $A$ results in collapse of the state vector
$\mid\Psi\rangle$ on one of the eigenvectors $\mid a\rangle$ of $A$
with eigenvalue $a$. Since, Born rule assigns the probability
$\mid\langle a\mid\Psi\rangle\mid^2$ for this event, the mean or the
expectation value is $\langle A\rangle=\sum_a a\mid\langle
a\mid\Psi\rangle\mid^2$. Interestingly, by noting that $a=\langle
a\mid A\mid\Psi\rangle/\langle a\mid \Psi\rangle$, the same
expression can be written in terms of an arbitrary basis rather than
$\{\mid a\rangle\}$. For example, for a hermitian operator $B$ with
eigenvectors $\mid b\rangle$,
\begin{equation}
\langle A\rangle=\sum_b\frac{\langle b\mid A\mid
\Psi\rangle}{\langle b\mid\Psi\rangle}\mid\langle
b\mid\Psi\rangle\mid^2 \label{weak}
\end{equation}
suggesting that $\langle b\mid A\mid \Psi\rangle/\langle
b\mid\Psi\rangle$, which is nothing but the weak value for
pre-selected $\mid\Psi\rangle$ and post-selected $\mid b\rangle$
states, is analogues to eigenvalue $a$ \cite{Aharonov3, Tollaksen}.
If we relax the rule that measurement of $A$ results in a state
vector in the set $\{\mid a\rangle\}$, we can generalize the
measurement axiom to include complex outcomes or the weak values.
For example, let us consider a spin-1/2 particle in spin up state
along $z$ direction. If we measure its spin along $x$ direction
($A=\sigma_x$), we find $+1$ or $-1$ with probabilities 1/2 which
gives zero mean value. If we consider the same measurement with
$B=\sigma_y$, corresponding values are, with equal probabilities,
$-i$ and $i$ which again results in vanishing expectation value.
Equation (\ref{weak}) suggests that $A_w$ can be replaced by its
real part. A more careful treatment of the problem by considering a
system weakly interacting with a measurement device shows that the
pointer of the latter is displaced by an amount given by the real
part of the weak value \cite{Aharonov}. It is also possible to
interpret a weak measurement as a positive operator valued measure
where the final state is not necessarily an eigenstate of the
measured dynamical value \cite{Jauch}.

Interpretation of complex values has been an important problem since
the early days of quantum mechanics. Obtaining probabilities by
evaluating modulus squares of complex probability amplitudes is the
most well known example. In this work, we present an alternative
interpretation of complex weak values by using their polar forms.

When $A$ is a projection operator so that it is of the form
$A=\mid\Omega\rangle\langle\Omega\mid$, after multiplying the
numerator and the denominator by $\langle\Psi\mid\Phi\rangle$, the
weak value becomes
\begin{equation}
A_w=\frac{\langle\Phi\mid\Omega\rangle\langle\Omega\mid
\Psi\rangle\langle\Psi\mid\Phi\rangle}{\mid\langle\Phi\mid\Psi\rangle\mid^2}=
\frac{\Delta_3(\Phi,\Omega,\Psi)}{\mid\langle\Phi\mid\Psi\rangle\mid^2}.
\end{equation}
Multiplication of the numerator and the denominator of $A_w$ by
$\langle\Psi\mid\Phi\rangle$ is a trivial mathematical operation but
the resulting expression has an important physical interpretation.
The cyclic multiplication of inner products of state vectors
$\Delta_3(\Phi,\Omega,\Psi)$ in the numerator of $A_w$ is a typical
example for the multi-vertex invariants introduced by Bargmann in
his alternative proof of the Wigner theorem and they are known as
Bargamann invariants \cite{Bargmann}. It is clear that Bargmann
invariants do not change under unitary operations while they are
mapped to their complex conjugates by anti-unitary transformations.
After Berry's discovery, it has been understood that the argument of
Bargamann invariant has a physical significance which is now
commonly known as geometric phase. We note that the same definition
of phase relation between nonorthogonal states has been introduced
by Pancharatnam in optics \cite{Pancharatnam}. Possible relation
between Bargmann invariant and Pancharatnam phases for the special
case of two state systems has been studied \cite{Tamate}.
Pancharatnam's geometric phase has been associated with the phase of
complex weak values \cite{Sjoqvist}. Geometric phase interpretation
of arguments of complex values in a quantum eraser interference
experiment has also been discussed \cite{Cormann}.

It has been shown that the argument of the Bargmann invariant, like
$\Delta_3(\Phi,\Omega,\Psi)$ in the numerator of $A_w$, is negative
of the geometric phase corresponding to the three geodesics
connecting the three state vectors \cite{Mukunda, Hangan}.
Therefore,
\begin{equation}
\arg A_w=\arg\Delta_3(\Phi,\Omega,\Psi)=-\phi_g(\Phi,\Omega,\Psi)
\end{equation}
where $\phi_g(\Phi,\Omega,\Psi)$ denotes the geometric phase
corresponding to the geodesic triangle with vertices
$\mid\Phi\rangle$, $\mid\Omega\rangle$, and $\mid\Psi\rangle$. More
generally, the phase of an $n-$vertex Bargmann invariant is
identical to the geometric phase for a closed ray-space curve
connecting the $n$ state vectors with geodesics. Since, n-vertex
structures can be triangularized, three-vertex invariants can be
thought of as the primitive building blocks of the geometric phase.
It is also possible to interpret this observation as another
manifestation of the fact that weak values respect sums, but not
products \cite{Hall}. We note that exchange of the pre- and the
post-selected states results in complex conjugation of the weak
value. Hence, time-reversal operation leads to simply a change in
sign of Berry's phase.

When an operator $A$ acts on a state vector $\mid\Psi\rangle$, the
resulting state can be decomposed into two pieces
\begin{equation}
A\mid\Psi\rangle=\langle A\rangle\mid\Psi\rangle+\Delta
A\mid\Psi^{\bot}\rangle
\end{equation}
where $\Delta A^{2}=\langle A^{2}\rangle-\langle A\rangle^{2}$ and
$\mid\Psi^{\bot}\rangle$ is a state vector perpendicular to
$\mid\Psi\rangle$ \cite{Aharonov4}. Therefore, the weak value can be
written as
\begin{equation}
A_w=\langle A\rangle+\Delta
A\frac{\langle\Phi\mid\Psi^{\bot}\rangle}{\langle\Phi\mid\Psi\rangle}
\end{equation}
and hence imaginary part of $A_w$ is dis proportional to the ratio
${\langle\Phi\mid\Psi^{\bot}\rangle}/{\langle\Phi\mid\Psi\rangle}$.
We note the the modulus of this ratio is nothing but $(1-p)/p$ where
$p=\mid\langle\Phi\mid\Psi\rangle\mid^{2}$ is the probability of
obtaining post-selected state $\mid\Phi\rangle$ for pre-selected
state $\mid\Psi\rangle$.

The idea of positive operator valued measures is quite old
\cite{Jauch}. However, the symmetric and informationally complete
case has been a subject of great interest because of the recent
developments in quantum information theory \cite{Renes,Zauner}.
Density matrix of a $d-$dimensional system has $d^2-1$ free
parameters. One of the main motivations of SIC-POVMs is to determine
these parameters in an optimal way. This task is achieved by
entangling the $d-$dimensional system with a $d^2-$dimensional one
so that projective measurements performed on the latter leads to
collapse of the former on equiangular states, \emph{i.e.} state
vectors having the same inner product moduli. In other words, by
using SIC-POVMs, arbitrary states of $d-$dimensional system are
collapsed on one of the $d^2$ equiangular state vectors in contrast
to standard projective measurements where only $d$ orthogonal final
states are possible. The completeness relation imply that
equiangular vectors $\mid\psi_i\rangle$ must satisfy
$\mid\langle\psi_i\mid\psi_j\rangle\mid=1/\sqrt{d+1}$ for $i\neq j$.
In 1999, Zauner conjectured that it is always possible to find $d^2$
such vectors in $d$ dimension and this is still an open problem
\cite{Zauner}. Introducing the projection operators
$\Pi_i=\mid\psi_i\rangle\langle\psi_i\mid$, we can write the overlap
condition as $\mathrm{Tr} \Pi_i\Pi_j=1/(d+1)$.

Now, let us consider a weak measurement where pre- and post-selected
states are $\mid\Psi\rangle=\mid\psi_i\rangle$ and
$\mid\Phi\rangle=\mid\psi_k\rangle$, respectively. Let us further
assume that
$A=\mid\Omega\rangle\langle\Omega\mid=\mid\psi_j\rangle\langle\psi_j\mid$,
Since, all of the inner products have the same modulus, the weak
value takes the form
\begin{equation}
A_w=\frac{e^{i\theta_{ijk}}}{\sqrt{d+1}}
\end{equation}
where $\theta_{ijk}\in[0,2\pi)$. We note that while the initial
$\mid\Psi\rangle=\mid\psi_i\rangle$ and the final
$\mid\Phi\rangle=\mid\psi_k\rangle$ states are fixed by projective
(strong) measurements, $\mid\psi_j\rangle$ is determined by a weak
measurement. In practice, SIC-POVMs can be applied on the system
with strong-weak-strong measurement pattern and weak values
corresponding to triples with different $i$, $j$ and $k$ values can
be used to evaluate $\theta_{ijk}$.

The simplest case to apply the suggested measurement procedure is
two level systems where $\mathrm{Tr} \Pi_i\Pi_j=1/3$ for any pair of
4 SIC-POVM vectors forming the vertices of a regular tetrahedron in
the Bloch sphere. Since there are 4 possible triplets, one might
expect 4 different $\theta_{ijk}$ values. However, symmetry of the
tetrahedron results in, apart from sign changes coming from
permutations, a single value which is $\theta_{ijk}=\pi/2$. This
special case is interesting in that all weak values are purely
imaginary. Therefore, measuring device reads a null value. In three
dimensions, it can be shown that 84 triple combinations (28 choices)
results in 5 different values \cite{Zhu}.

For SIC-POVMs, Bargmann invariants
$\mathrm{Tr}\Pi_i\Pi_j\Pi_k=e^{i\theta_{ijk}}/(d+1)^{3/2}$ become
completely symmetric 3-index tensors and they play a central role in
representation of density matrices. A density operator
$\rho=\sum_i\lambda_i\Pi_i$ describes a pure state if and only if
$\mathrm{Tr} \rho^2=\mathrm{Tr} \rho^3=1$ \cite{Fuchs}. Therefore
along with $\Sigma_i\lambda_i=\Sigma_i\lambda_i^2=1$,the equation
\begin{equation}
\Sigma_i\lambda_i^3+\frac{6}{d\sqrt{d+1}}\Sigma_{\{i,j,k\}}\cos\theta_{\{i,j,k\}}\lambda_i\lambda_j\lambda_k=1
\end{equation}
determines all the valid expansion coefficient sets $\{\lambda_i\}$
corresponding to a pure state. Here, the sum is over triple
combinations $\{i,j,k\}$ where all indices are different from each
other. Since, we are summing over combinations, $\{i,j,k\}$ and
$\{j,i,k\}$ denote the same term. Six possible permutations
contribute only two different values (depending on wheter the
permutation is even or odd) and therefore we can combine them in a
single cosine term. Clearly, the angles $\theta_{\{i,j,k\}}$ are not
independent. The relations among $\theta_{ijk}$'s restrict
$\theta_{\{i,j,k\}}$'s, too. For example,
$\theta_{123}\theta_{134}\theta_{124}=\theta_{234}$ implies a
similar relation for $\theta_{\{i,j,k\}}$'s. Obtaining the
corresponding probabilities $p_i$ from the expansion coefficients
$\lambda_i$ is straightforward from the relation
$\lambda_i=(d+1)p_i-1/d$ \cite{Fuchs}.

For an arbitrary projection operator
$A=\mid\Omega\rangle\langle\Omega\mid$, we can ask if the weak value
$A_w$ is real. This is equivalent to looking for vanishing Bargmann
invariant phase or geometric phase. Null phase curves have been
introduced in search of a generalized connection between Bargmann
invariants and geometric phases \cite{Rabei}. Geometric phase
vanishes when three vertices lie on a null phase curve. This is also
the case when the curve is a geodesic. Null phase condition is
necessary for geodesics but not sufficient. It is possible to
generalize this definition to arbitrary phase values. In this case,
each constant phase curve connecting the pre- and the post-selected
states will describe the projection operators giving that particular
weak value argument.

In conclusion, arguments of quantum weak values can be related to
geometric phases in a very simple way. Therefore, not only their
real and imaginary parts but also moduli and arguments of weak
values have a physical significance, the latter being a direct
measure of Berry's phase. In particular, alternating projective and
weak SIC-POVMs can be used to determine the angular relations among
projection operators. For a projection operator, the question of
when a weak value is a real number turns out to be equivalent to
having real Bargmann invariants.

The author acknowledges helpful discussions with G. B. Ba\u{g}c{\i},
O. Pusuluk, and V. Vedral.

\end{document}